\documentclass[prb,twocolumn,groupedaddress]{revtex4}

\usepackage[dvips]{color}
\usepackage{graphicx}
\usepackage {amssymb}

\begin{document}

\title{Integration of a gate electrode into carbon nanotube devices for scanning tunneling microscopy }

\author{J. Kong}
\author{B.J. LeRoy}
\author{S.G. Lemay}
\author{C. Dekker}
\email[Electronic mail: ]{dekker@mb.tn.tudelft.nl}
\affiliation{Kavli Institute of Nanoscience, Delft University of
Technology, Lorentzweg 1, 2628 CJ Delft, The Netherlands}

\date{\today}

\begin{abstract}
We have developed a fabrication process for incorporating a gate
electrode into suspended single-walled carbon nanotube structures
for scanning tunneling spectroscopy studies.  The nanotubes are
synthesized by chemical vapor deposition directly on a metal
surface.  The high temperature (800~$^{\circ}$C) involved in the
growth process poses challenging issues such as surface roughness
and integrity of the structure which are addressed in this work.
We demonstrate the effectiveness of the gate on the freestanding
part of the nanotubes by performing tunneling spectroscopy that
reveals Coulomb blockade diamonds. Our approach enables combined
scanning tunneling microscopy and gated electron transport
investigations of carbon nanotubes.
\end{abstract}

\maketitle

Single-walled carbon nanotubes (SWCNTs) have attracted tremendous
attention in recent years due to their unique properties and
promising applications.~\cite{1,2} In particular, their electronic
properties have been under intensive study, both for interest in
fundamental one-dimensional physics and for exploration of
potential molecular electronic devices.~\cite{3,4} Scanning
tunneling microscopy (STM) is a powerful tool for characterizing
SWCNTs since the atomic and electronic structures can be obtained
simultaneously for each nanotube.~\cite{5,6}  However, the close
proximity of a conducting substrate and the difficulty of
incorporating a gate electrode~\cite{leoscangate} limits the range
of phenomena that can be studied using STM. Both of these
limitations can be overcome by suspending the SWCNT over a trench
with a gate electrode at the bottom of the trench (Fig 1(a)).  In
such a device geometry, electron transport measurements can be
combined with STM studies. Even without the gate electrode or
separate source and drain contacts, new effects are observed using
the STM over the freestanding portion of the
SWCNT.~\cite{LeRoyAPL,LeRoyNature} The addition of a gate
electrode allows addressing a wealth of extra information such as
band bending along the nanotube, the electronic structure during
transport, electromechanical effect upon gating, etc.

In this Letter, we demonstrate the fabrication of a suspended
SWCNT device with an integrated gate electrode for STM
investigation. The conventional method of nanotube sample
preparation for STM involves sonicating SWCNTs in organic solvents
followed by depositing the suspension over a conducting substrate.
When nanotubes are deposited over a surface with trenches we find
that they usually sag over the trench (data not shown), presumably
due to capillary forces upon drying of the solvent.  The sagging
of the SWCNT significantly hinders access to the suspended portion
of the SWCNT. To overcome this difficulty, we chose to directly
grow the nanotubes on a conducting surface by chemical vapor
deposition (CVD). This has the additional advantage of allowing us
to image the as-grown nanotubes without any post-processing. The
main challenge in realizing such devices comes from the high
temperature involved in the CVD process.  This brings up issues
such as choice of substrate material, surface roughness and device
integrity. We will address these in the following and show that it
is possible to realize a 3-terminal device, with the capability of
gating and simultaneous atomic-resolution imaging and
spectroscopy.

\begin{figure}[ht]
\includegraphics[width=3.35in]{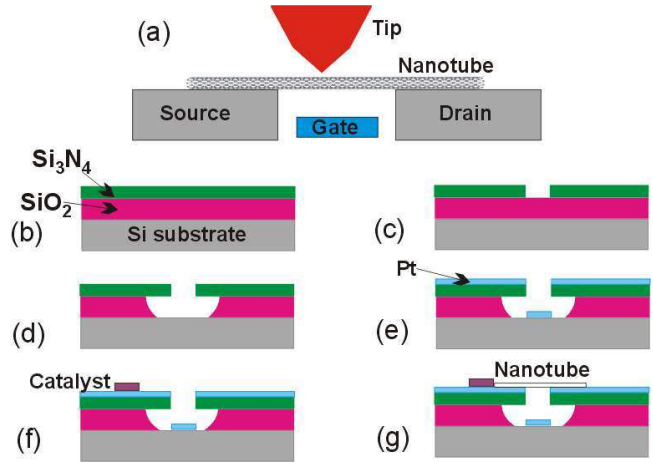}
\caption{(Color online) (a) Diagram of suspended nanotube
structure for combined electron transport and STM studies. (b)-(g)
Schematic of the fabrication process.  (b) Starting substrate: Si
with 500 nm thermally grown SiO$_2$ and 100 nm Si$_3$N$_4$.  (c)
Patterning and dry etching of trenches in Si$_3$N$_4$ layer.  (d)
Wet etching of SiO$_2$.  (e) Evaporation of Pt (with 1 nm Ti as
sticking layer) (f) Patterning and catalyst deposition.  (g) After
CVD growth, nanotubes are suspended over the trenches.}
\end{figure}

We choose Pt as the material for the conducting substrate.  It is
a noble metal which does not develop oxides in air, and is
therefore suitable for imaging with STM.  The melting point of Pt
is 1772~$^{\circ}$C, high enough to be compatible with the CVD
process.  In addition, it does not form hydrides at elevated
temperatures~\cite{11} (the CVD process takes place in a hydrogen
rich environment).  The Pt-carbon phase diagram shows a eutectic
alloying temperature at 1705$\pm$13~$^{\circ}$C,~\cite{12} much
higher than the CVD temperature.  Consequently the presence of Pt
does not interfere with the carbon nanotube
growth.~\cite{GolovchenkoAPL,HongjieDoublegate}

Our process starts with heavily doped p++ Si substrates with a 500
nm thermally grown oxide and 100 nm of Si$_3$N$_4$ on top (Fig
1(b)). Trenches in Si$_3$N$_4$ are opened by electron beam
lithography (EBL) and anisotropic reactive ion etching with
CHF$_{3}$ and O$_{2}$. Subsequently the resist is removed by
acetone and the surface is cleaned by O$_{2}$ plasma (Fig 1(c)).
An isotropic wet etch (Buffered HF) deepens the trench into the
SiO$_{2}$ and creates an undercut (Fig 1(d)). This is followed by
direct electron-beam evaporation of Pt over the whole substrate
(Fig 1(e)), with 1 nm Ti as sticking layer. The undercut prevents
the metal on the top from contacting the bottom. The Pt at the
bottom of the trench, which forms PtSi after heating, is used as
the gate electrode~\cite{note2}. The positions for the catalyst
are defined by EBL, and an alumina-supported iron catalyst is
deposited on the substrate followed by lift-off in
acetone~\cite{KongNature} (Fig 1(f)). As a final step, nanotubes
are grown by CVD at 800~$^{\circ}$C (Fig 1(g)). The synthesis is
carried out in a 1 inch diameter tube furnace under the flow of
910 mL/min of CH$_{4}$, 700 mL/min of H$_{2}$ and 40 mL/min of
C$_{2}$H$_{4}$~\cite{14} for 5 minutes.

\begin{figure}[ht!]
\includegraphics[angle=0,width=3.35in]{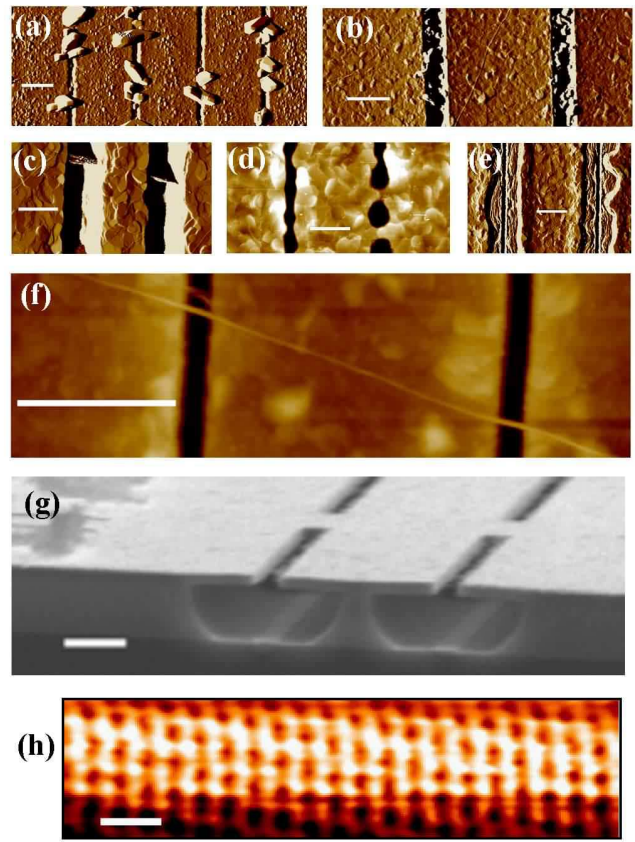}
\caption{(Color online) (a) AFM image of faceted Pt particles
decorating the trench edges when heating the substrate in Ar.  (b)
AFM image of the device when H$_{2}$ is used instead of Ar but
there is still a trace amount O$_{2}$ in the system.  (c)-(e) AFM
images showing trench widening and corrugation due to the heating.
(f) AFM image of the result using the fast heating method.  (g)
cross-sectional SEM image of the final structure.  (h) STM atomic
resolution topography image on a suspended part of a SWCNT. The
image is taken with a sample voltage of -0.1 V and a feedback
current of 300 pA. Scale bars: (a)-(g): 500~nm, (h): 5 \AA}
\end{figure}

We have found that the final Pt surface depends very sensitively
on trace amounts of O$_{2}$ present during the CVD process.  The
CVD synthesis takes place at 800~$^{\circ}$C.  When heating up and
cooling down the reactor, a constant flow of H$_2$ (purity
99.999\%) is applied.  This step is crucial to obtain the desired
trench structure for STM imaging.  Fig 2(a) displays the result
when Ar (purity 99.996\%) instead of H$_2$ was used during heating
up. Particles with crystalline facets can be seen decorating the
edge of the trenches.  These particles appear even when using a
substrate without catalyst, and we conclude that they are Pt
crystallites. It has been recognized in electrochemical studies
that the oxidizing-reducing cycle leads to Pt surface
reconstruction.~\cite{16} This was attributed to the fact that the
oxidized Pt species diffuses across the surface more easily.
Presumably in our case, Pt atoms are oxidized by the trace amount
of O$_2$ ($<$10 ppm~\cite{17}) in the Ar and accumulate along the
trench edges. Later during the CVD process they are reduced again
in the H$_2$ rich environment. This leads to the faceted Pt
particles decorating the trenches.  In contrast, when H$_2$ is
used during the heating up, it reduces the O$_2$ impurities,
therefore significantly lessening the effect as shown in Fig 2(b).
In order to even further reduce exposure to O$_2$, we flush the
system at room temperature with H$_2$ for 20 min. As a result, the
Pt migration is effectively eliminated (Fig~2(f)).

Fig 2(c) - (e) display another issue we encounter during sample
fabrication, i.e, the trenches become distorted after CVD
synthesis. The effect becomes more severe when the thickness of
the Pt film decreases.  The Pt in Fig 2(c) is 100 nm thick and the
trenches are 300 nm wide before the CVD. After the high
temperature they appear to have become $\sim$ 450 nm wide.
Although the trenches are widened, the edges are still straight
and sharp. When the thickness of the Pt is decreased to 50 nm, the
edges of the trenches become noticeably rougher as shown in Fig
2(d). Occasionally, this also causes short-circuiting of the two
sides of the trench. When the Pt film thickness is reduced to 20
nm, the widening effect becomes universal. Fig 2(e) displays the
edges of such a thin Pt film retracting away from 30 nm wide
trenches.  As suggested by this figure, this effect is probably
due to the surface tension between the Pt film and the substrate
at the edge of the trench, causing the Pt to be dragged away.  The
migration of the Pt film causes a severe problem for the integrity
of the trench structure.

In order to prevent the degradation of devices, we adopted a fast
heating/cooling technique.~\cite{18}  The furnace was first heated
up to the target temperature (800~$^{\circ}$C) without the
substrate. Once the temperature became stable, the desired gases
were switched on and the substrate was quickly brought to the
center of the furnace. After CVD synthesis for 5 min, the
substrate was immediately brought out of the high temperature
region.  In this way the duration of substrate heating and the
deformation of the Pt film were minimized. Fig 2(f) is a
representative AFM image of the devices fabricated by this fast
heating/cooling technique.  Fig 2(g) is a cross-sectional SEM
image of the device after CVD. Fig 2(h) presents a STM topography
image of an individual metallic SWCNT suspended across a 200 nm
trench demonstrating atomic resolution.

\begin{figure}[ht]
\includegraphics[angle=0,width=3.35in]{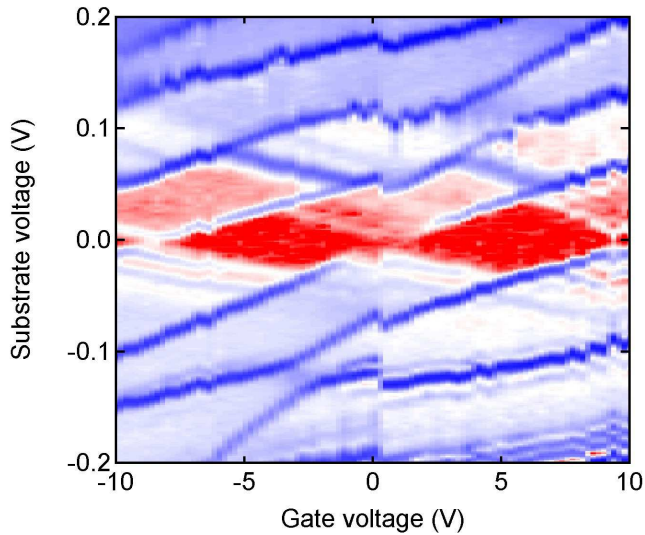}
\caption{(Color) Logarithm of differential conductance as a
function of substrate and gate voltage for the nanotube in Fig
2(h) suspended over a 200 nm wide trench. The STM tip is held at a
fixed location above the suspended portion of the SWCNT, with a
setpoint current of 500 pA at a substrate voltage of -0.5 V with
respect to the tip. Red corresponds to low values and blue
corresponds to high values.}
\end{figure}

In order to demonstrate the potential of these devices, we present
gated STM measurements.  Our previous studies on non-gated
nanotubes showed that the suspended parts of the nanotubes tend to
form quantum dots and that Coulomb peaks are observed in the
differential conductance dI/dV.~\cite{LeRoyAPL} With the metal
electrode at the bottom of the trench as a gate, the
single-electron-charging behavior can be investigated in full
detail. For the same nanotube imaged in Fig 2(h), we have
performed tunneling spectroscopy measurements as a function of
gate voltage. We used the STM tip as the source, the substrate as
the drain, and the bottom electrode as the gate. The measurements
were performed by adding a small ac voltage to the substrate
voltage and using lockin detection at a temperature of 5 K.

The measured dI/dV as a function of sample and gate voltage is
plotted in Fig. 3. It shows the characteristics of a typical
Coulomb diamond plot.~\cite{19} Near zero substrate voltage, there
are diamond shaped regions where the current is blocked and at
higher voltages there are peaks associated with adding additional
electrons to the SWCNT.  Single-electron tunneling is observed
with a Coulomb charging energy of about 40 meV.  The periodicity
of the diamonds and the lack of additional peaks implies that we
are probing a single quantum dot. We do not see any evidence for
multiple quantum dots in series or parallel such as, for example,
induced by defects in the SWCNT. The ability to access the full
Coulomb diamond plot allows us to unambiguously identify the
origin of all the peaks in the differential conductance.  The
peaks sloping from the bottom left to the top right of the image
are due to the Fermi energy of the substrate lining up with a
state on the SWCNT. The peaks running from the top left to the
bottom right are due to the Fermi energy of the tip lining up with
states in the SWCNT. Excitations lines can be observed parallel to
the edges of the Coulomb diamonds, and side peaks due to
phonon-assisted tunnelling~\cite{LeRoyNature} are also visible.

The addition of the gate electrode allows the values of the
capacitances in our device to be determined.  From the slope of
the edges of the Coulomb diamonds we can derive the capacitances:
C$_{tip}$ $\approx$ 1.8 aF, C$_{substrate}$ $\approx$ 2.5 aF.  The
width of the diamond in gate voltage gives the value for
C$_{gate}$ $\approx$ 0.018 aF. These values are in good agreement
with a numerical simulation, where the tip is modelled as a 150 nm
radius sphere.  Experimentally, we find that the value of
C$_{substrate}$ can vary between 1 and 10 aF depending on the
extent of the quantum dot over the Pt substrate. The small
SWCNT-substrate capacitance and hence large charging energy
implies that the quantum dot does not extend very far over the Pt
substrate.  The gate coupling is relatively weak in this structure
because of the presence of the Pt substrate and the STM tip. These
large metal surfaces act to shield the gate from the SWCNT
reducing its coupling.

In summary, we have incorporated a gate electrode into devices for
STM studies of SWCNTs. Coulomb diamond plots were obtained using
STM with simultaneous atomic-resolution imaging capability. With
proper design, the metal on the two sides of the trench can be
isolated and function as independent source and drain electrodes
of the nanotube device. Combining transport measurements with STM
studies of suspended SWCNTs opens many opportunities for further
studying the electronic and mechanical properties of nanotubes.

The authors would like to thank FOM and NWO for funding, and A.
Hassanien and J.-O Lee who were involved in earlier stages of this
work.










\begin{thebibliography} {99}
\bibitem{note2} Alternatively, we can etch the SiO$_2$ halfway, leaving the Pt inside the trench floating.  The p++ Si then serves as the gate.
\end{thebibliography}
\end{document}